\documentclass[preprint, double-spaced]{revtex4}
\usepackage[colorlinks=true,linktocpage=true,linkcolor=blue,citecolor=blue]{hyperref}
\usepackage{graphicx}
\usepackage{amsmath,bbm}
\usepackage{amssymb,bm}
\usepackage{array}

\begin{document}
\title{Study of charged particle production in $U$-$U$ collisions in the Wounded Quark Model}
\author{O. S. K. Chaturvedi$^1$}
\author{P. K. Srivastava$^{2,}$\footnote{prasu111@gmail.com, prashant.srivastava@iitrpr.ac.in}}
\author{Ashwini Kumar$^1$}
\author{B. K. Singh$^{1,}$\footnote{bksingh@bhu.ac.in}}
\affiliation{$^1$Department of Physics, Institute of Science, Banaras Hindu University, Varanasi-221005, India}
\affiliation{$^2$Department of Physics, Indian Institute of Technology Ropar, Rupnagar - 140001, India}
\begin{abstract}
Recently, there has been a growing interest in the study of deformed uranium-uranium ($U$-$U$) collisions in its various geometrical configurations due to their usefulness in understanding the different aspects of quantum chromodynamics (QCD). In this paper we have studied the particle production in deformed $U$-$U$ collisions at $\sqrt{s_{NN}}$ = $193$ GeV using modified wounded quark model (WQM). At first, we have shown the variation of quark-nucleus inelastic scattering cross-section ($\sigma_{qA}^{in}$) with respect to centralities for various geometrical orientations of $U$-$U$ collisions in WQM. After that we have calculated the pseudorapidity density ($dn_{ch}/d\eta$) within WQM using two-component prescription. Further we have calculated the transverse energy density distribution ($dE_{T}/d\eta$) along with the ratio of transverse energy to charged hadron multiplicity ($E_{T}/N_{ch}$) for $U$-$U$ collisions and compared them with the corresponding experimental data. We have shown the scaling behavior of $dn_{ch}/d\eta$ and $dE_{T}/d\eta$ for different initial geometry of $U$-$U$ collision with respect to $p$-$p$ data at $\sqrt{s_{NN}}=200$ GeV. Furthermore we have shown the Bjorken energy density achieved in $U$-$U$ collisions for various configurations and compared them with experimental data of $Au$-$Au$ at 200 GeV. We observe that the present model suitably describes the experimental data for minimum bias geometrical configuration of $U$-$U$ collisions. An estimate for various observables in different initial geometries of $U$-$U$ collisions is also presented which will be tested in future by experimental data.
\end{abstract}

\maketitle 
\section{Introduction}
\noindent
The main endeavor of heavy-ion collision experiments is to study the nature of quantum chromodynamics (QCD) at extreme physical conditions. In these experiments, variety of colliding species has been used at various beam energies to understand the different aspects of QCD medium ~\cite{Singh_Reports:0993,SinghIJMA:1992,Dremin:2001}. For example hadron-hadron collisions provide us the basic scattering cross-sections for various charged hadron productions. Hadron-nucleus interactions quantify the nuclear effects on this production cross-sections and finally nucleus-nucleus collisions can shed light on the effect of various phases of QCD matter~\cite{QPM:2010} on the particle production processes. Further, the change in collision energy facilitate the study of strongly interacting matter in various physical conditions e.g., at high temperature and zero chemical potential or at high chemical potential and zero temperature or at moderate temperature and chemical potential etc~\cite{QPM:2010}. Even in nucleus-nucleus collisions, various species have been used to study other properties of QCD and related phases. For example, $Cu$-$Au$ collision is a good tool to study the electromagnetic and chiral magnetic behavior of QCD matter due to a sizable amount of directed flow arises in these collisions ~\cite{Hirono:2014}. $Au$-$Au$ and $Pb$-$Pb$ collisions are useful to study the phase transitions and the behavior of exotic phases of strongly interacting matter. In recent years collision among uranium nucleus has gained a lot of interest because of their non-spherical geometry which is actually prolate ~\cite{Ulrich:2005,Kuhlman:2005}. The deformed nuclei like $U^{238}$ promise additional gain of energy density compared to spherical nuclei at relativistic energies ~\cite{Ulrich:2005,Kuhlman:2005,Jakob:1976}. The properties of QGP characterized by the observables like elliptic flow, jet quenching, charmonium suppression and multiplicity can be better understood by the collision of deformed uranium nuclei due to its initial geometry and specific orientation ~\cite{Hiroshi:2009,Shou:2015,Sergei:2010}.  The collision of uranium nuclei can provide a reliable tool for the detection of chiral magnetic effect (CME) ~\cite{Sergei:2010}. The main hurdle to detect the CME signal reliably is the background effect of same strength generated due to elliptic flow. In spherical nuclei, it is difficult to disentangle both these effect since the strength of both the signals generated from elliptic flow and CME is of similar strength in peripheral collisions. However, in $U$-$U$ central collisions, the different geometrical orientations can provide a way to subtract the background signal from CME signal due to a measurable difference in their strength. Thus central collisions of $U$-$U$ nuclei in tip-tip configuration can possibly be a good tool to characterize the signal of CME ~\cite{John:1311,Bjorn:1403}.

In case of central body-body collisions, the eccentricity is of the order of $0.25$, almost as large as in the semicentral $Au$-$Au$ at $b=7$ fm where the maximum density is much lower ~\cite{Ulrich:2005}. Therefore the body-body configurations of $U$-$U$ collisions allow us to study the elliptic flow with higher density at given beam energy ~\cite{Xiao:2007}. Dependence of energy loss of partons on path length and medium density is not yet fully understood since the fireball size created in $Au$-$Au$ collision is small and thus the difference in path length for the parton traversing in-plane and out-of-plane is small. Different geometrical configurations of $U$-$U$ collisions may provide almost twice as much difference between the in-plane and out-of-plane path lengths for the same eccentricities as semicentral $Au$-$Au$ collisions and thus provide a better opportunity to understand the energy loss mechanism ~\cite{Shuryak:9906062}. Deformed nuclei like uranium, having five independent parameters: impact parameter and four polar angles instead of one impact parameter as in the case of spherical nuclei, can provide in-depth knowledge of charged hadron production and collision dynamics of QCD matter ~\cite{Kuhlman:2005}. In the case of $J/\psi$, nuclear effects like shadowing, absorption etc., are crucial to disentangle the suppression caused by cold nuclear matter effects and by quark-gluon plasma (QGP). Different orientations of $U$-$U$ interactions could study nuclear effects with reduced uncertainty and provide an additional check on the models which successfully describing the $Au$-$Au$ data regarding $J/\psi$ ~\cite{Shuryak:9906062,Daniel:2011}. Further it has been suggested that central $U$-$U$ collision in the body-body configuration (a high baryon density system) represents an optimum system to study the quark-hadron phase transition in its beam energy dependence ~\cite{Peter:054909,PeterPLB:1999,Bao-An:021903,pfkolb}. 

A lot of effort has been taken to simulate and analyze the charged particle distributions in deformed $U$-$U$ collisions ranging from fixed target experiments e.g. super proton synchrotron (SPS) to collider experiment like relativistic heavy-ion collider (RHIC). Most of these models are based on wounded nucleon approach ~\cite{Ulrich:2005,Kuhlman:2005,Sergei:2010,John:1311,Bjorn:1403,Ulrich:2005,Xiao:2007,Shuryak:9906062,Kuhlman:2005,Daniel:2011,Peter:054909,Bao-An:021903,Wang:1406,Ke-Jun:2008,Tetsufumi:2011,Rihan:2012,Xu:2012,Rybczy:2013,Bjorn:2014,Sandeep:2015,Andy:2015,Piotr:2016,Drees:1312,Hoelck:1602,Henning:1999,Nepali:2007,Peter:2009,Sergei:2010,Moller:1995,Hagino:2006}. However, in this article, we study the particle production in a wounded quark approach which assumes quark-quark interaction as the basic entity in defining the nucleus-nucleus interaction. We have shown in our earlier publications that this phenomenological model can describes the multiplicity distribution data simultaneously for various types of symmetric and asymmetric collisions at all the centralities ranging for most-peripheral to most-central ~\cite{Ashwini_2:2013,Ashwini:2013,Chaturvedi:2016}. Thus, it is worthwhile to investigate whether a wounded quark approach can give reliable prediction of charged hadron production in deformed uranium nuclei collisions and provide the estimate for observables in various initial geometrical configurations. Here, we have extended our modified version of WQM ~\cite{Ashwini_2:2013,Ashwini:2013,Chaturvedi:2016,Shyam_1:1989,Shyam:1985,Singh_3:1986} with minimal number of parameters to explain and predict the charged hadron distributions with various controlling parameters in deformed $U$-$U$ collisions for each centrality at a colliding energy of $193$ GeV. The rest of the paper is organized as follows: In Section II, we have presented the description of two-component WQM using suitable nuclear charge density function for uranium nucleus along with description of wounded quarks and mean number of quark-quark collisions. Further, in section III we have shown the pseudorapidity density within WQM, their comparison with the corresponding experimental data, scaling with respect to $p$-$p$ data and estimate for energy density. Finally we have summarized our present analysis in section IV.

\section{Model Formalism}

In the heavy-ion collision experiments, temperature and energy density of QCD medium are two important quantities whose information can be extracted from pseudorapidity distribution of charged hadrons. In recent publications ~\cite{Ashwini_2:2013,Ashwini:2013,Chaturvedi:2016}, we have proposed a parametrization for the pseudorapidity distribution ($dn_{ch}/d\eta$) of charged hadrons produced in $p$-$p$ collisions based on the multiparton exchange as considered in additive quark model (AQM) ~\cite{BialasPRD:1982,Anisovich:1984,lipkin}. The basic theme of our model can be described in three points : (1) One, two or three gluons have been exchanged between a quark of the first nucleus or hadron with the quarks belonging to the other nucleus or hadron. (2) The resulting color flux tube or color string somewhat stretched between them and other constituent quarks because they try to restore their color singlet behavior and thus the energy in the color flux tube increases. (3) The color tubes thus formed finally break-up into new hadrons and/or quark-antiquark pairs. In the present model, we have accommodated new data coming from $p$-$p$ collisions at 0.9, 1.8, and 7 TeV energies~\cite{Adam_Alice} along with other $p-p$ data by providing a parametrization for $dn_{ch}/d\eta$ at mid-rapidity as follows:
\begin{equation}
 <(dn_{ch}/d\eta)^{pp}_{\eta=0}>=(a_{1}^{'}+b_{1}^{'} ln \sqrt{s_{a}}+c_{1}^{'}ln^{2} \sqrt{s_{a}})-{\alpha_1}'.
\end{equation}
We have obtained the values of the parameters as  $a_{1}' = 1.15$, $b_{1}' = 0.16$, and $c_{1}' = 0.05$~\cite{Chaturvedi:2016}. In Eq. (1), $\alpha$~\cite{Singh_3:1986,Shyam:1985,Ashwini:2013} is the leading particle effect which arises due to the energy carried away by the spectator quarks and its experimental value is determined as $0.85$. The first, second and third term in right hand side of above parametrization (Eq. (1)) are shown in Fig. 1. The first term which remains constant with respect to collision energy arises due to single gluon exchange between the wounded quark of target and projectile. Second term which depends on the logarithm of available collision energy ($\sqrt{s_{a}}$) arises due to two gluon exchange via pomeron contribution to scattering cross-section as proposed earlier by Nussinov~\cite{nussinov} and Low~\cite{low}. Similarly the third term in the parametrization is the contribution of three gluon exchange among the wounded quarks~\cite{alexopoulos,walker}. The coefficients of these three terms i.e., $a,~b$ and $c$ actually control the contribution of these three channels. 

Taking the assumption of additive quark model~\cite{BialasPRD:1982,Anisovich:1984,lipkin}, mean number of collisions of wounded quark inside the nucleus $A$ is defined as the ability of constituent quark in the projectile hadron to interact repeatedly inside a target nucleus and can be expressed as :
\begin{equation}
\nu_{qA} = \frac{A\sigma_{qN}^{in}}{\sigma_{qA}^{in}}.
\end{equation}
Here $\sigma_{qN}^{in}$ and $\sigma_{qA}^{in}$ are the inelastic cross-sections for quark-nucleon ($q$-$N$) and quark-nucleus ($q$-$A$) interactions, respectively. $\sigma_{qN}^{in}$ is calculated from the non-diffractive cross-section of nucleon-nucleon collision ($\sigma_{NN}^{ND}$) using the fundamental assumption of AQM i.e., $\sigma_{qN}^{in} = (1/3)\sigma_{NN}^{ND}$. $A$ is the atomic mass of the target nucleus.

The mean number of inelastically interacting quarks with the nuclear target $A$ can be written as :
\begin{equation}
N_{q}^{hA} = \frac{N_{c}\sigma_{qA}^{in}}{\sigma_{hA}^{in}}.
\end{equation}
Physically, $N_{q}^{hA}$ is equal to the number of color flux tube created between projectile and target. In Eq. (3), $N_{c}$ is the number of valence quarks in the hadron $h$. $\sigma_{qA}^{in}$ and $\sigma^{in}_{hA}$ are the scattering cross sections for quark-nucleus and hadron-nucleus interactions.

\begin{figure}
\includegraphics[scale=0.50]{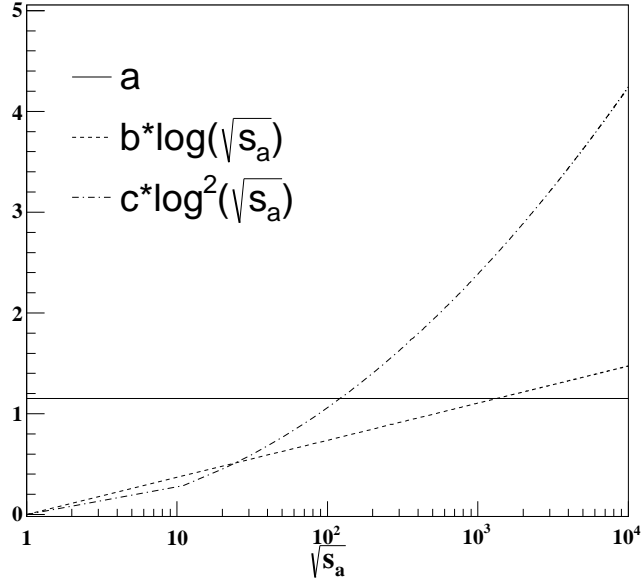}
\caption{Variation of each factor in the parametrization as given by Eq. (1) with respect to available collision energy ($\sqrt{s_{a}}$).}
\end{figure}
\begin{figure}
\includegraphics[scale=0.35]{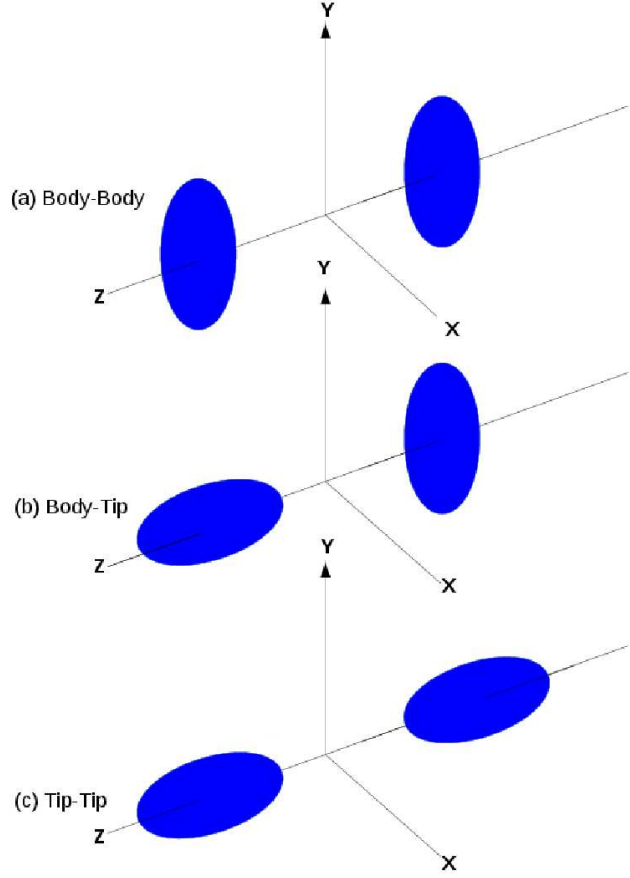}
\caption{Variation of various possible initial geometrical configuration in central $U$-$U$ collisions.}
\end{figure}

For deformed uranium nuclei we have used the modified form of Woods-Saxon nuclear density distribution ~\cite{Shou:2015,Loizides:1408} as follows: 
\begin{equation}\label{eq:deformed}
  \rho(r,\theta)=\rho_0 \frac{1}{1+\exp\frac{\left(r-R(1+\beta_2 Y_{20} +\beta_4 Y_{40})\right)}{a}}\,,
\end{equation} 
where $Y_{20}=\sqrt{\frac{5}{16\pi}}(3{\rm cos}^2(\theta)-1)$, 
$Y_{40}=\frac{3}{16\sqrt{\pi}}(35{\rm cos}^4(\theta)-30 {\rm cos}^2(\theta)+3)$ are the spherical harmonics with the
deformation parameters $\beta_2$ and $\beta_4$. 
The different parameters value for uranium nuclei is taken from Refs. ~\cite{Loizides:1408,Shou:2015}

The profile function $D_{A}(b)$ is related to nuclear density, $\rho(r,\theta)$ by the relation
\begin{equation}
D_{A}(b) = \sum_{\theta}\int_{-\infty}^{\infty}\int_{0}^{2\pi}\rho(r,\theta)dz d\phi,
\end{equation}
as $r$ is related to $b$ and $z$ by the following relation,
\begin{equation}
r= \sqrt{b^{2}+z^{2}}.
\end{equation}
For tip and body orientations of nucleus, we take a particular value of $\theta$ as $0$ and $\pi/2$, respectively. For minimum-bias, we take sum over polar angle $\theta$. Further we integrate over $\phi$ from $0$ to $2\pi$ and thus do not choose any special orientation of nucleus in $\phi$-space. The quark-nucleus inelastic interaction cross-section $\sigma_{qA}^{in}$ is determined from $\sigma_{qN}^{in}$ by using Glauber's approximation (neglects the Glauber series all excited states and includes only the ground states of the colliding objects) as follows :
\begin{equation}
\sigma_{qA}^{in}=\int d^{2}b\left[1-\left(1-\sigma_{qN}^{in}D_{A}(b)\right)^{A}\right],
\end{equation}

Now we move towards the main theme of our paper: the extrapolation of WQM from hadron-nucleus collision to nucleus-nucleus collisions. All the elements of hadron-nucleus collisions are present and same for nucleus-nucleus collisions. However, the number of participating quarks, number of color strings formed and the mean number of collisions are quite large for nucleus-nucleus collisions. The mean number of collisions happens for a wounded (participating) quark can be calculated from the following expression which is nothing but the multiplication of mean number of collisions for a wounded quark in $A$ within nucleus $B$ and the mean number of collisions for a wounded quark in $B$ within nucleus $A$ : 
\begin{equation}
\nu_{q}^{AB}=\nu_{qA}\nu_{qB}=\frac{A\sigma_{qN}^{in}}{{\sigma_{qA}^{in}}}.\frac{B\sigma_{qN}^{in}}{{\sigma_{qB}^{in}}}.
\end{equation}
Furthermore, the mean number of participating quarks $N^{AB}_{q}$ (or in sort denoted by $N_{q}$) can be calculated by Eq. (9) in the following manner:
\begin{equation}
N^{AB}_{q}=\frac{1}{2}\left[\frac{N_{B}\sigma_{qA}^{in}}{{\sigma_{AB}^{in}}}+\frac{N_{A}\sigma_{qB}^{in}}{{\sigma_{AB}^{in}}}\right],
\end{equation}
where $\sigma_{AB}^{in}$ is the inelastic cross-section for $A$-$B$ collision. To calculate $\sigma_{AB}$ we take the help of optical model as discussed in Refs.~\cite{fernbach,hoang} and can be expressed in the following manner:
\begin{equation}
\sigma_{AB}^{in}= \pi r^{2}\left[A^{1/3}+B^{1/3}-\frac{c}{A^{1/3}+B^{1/3}}\right]^2.
\end{equation}

In the above expression, the last term in the bracket of right hand side resembles the ``overlapping parameter'' of Bradt-Peters formula for scattering cross-section. The constant $c$ is related with the mean free path of a nucleon inside a nucleus and has a value $4.45$ for nucleus-nucleus collisions.

Wounded quark model which is based on additive quark model (AQM)~\cite{BialasPRD:1982,Anisovich:1984,lipkin} assumes particle production in hadron-hadron collisions due to three different sources : the central rapidity region populates from the contribution of breaking of a colored string (the produced particles scales with number of participating quarks); the fragmentation of quarks which exchanged a colored gluon populates the fragmentation regions of the projectile and the target (scales with mean number of quark collisions); the fragmented quarks in spectator region which did not exchange gluons populates also in the projectile and target fragmentation regions. Thus the production mechanism in different rapidity regions are quite separated from each other in hadron-hadron collisions. However in the case of nucleus-nucleus collisions there are still some uncertainties in construction of produced particles as there is overlap of various mechanism whose contribution is not well known. Thus the multiplicity production in central rapidity region may have contribution from breaking of colored strings as well as from the fragmentation of quarks.  Consequently the multiplicity in mid-rapidity region might not scale with number of wounded quarks exactly. Based on these arguments we proposed a two component WQM ~\cite{Singh_3:1986,Shyam:1985,Ashwini:2013}: 
\begin{equation}
\left(\frac{dn_{ch}}{d\eta}\right)^{AA}_{\eta=0}=\left(\frac{dn_{ch}}{d\eta}\right)^{pp}_{\eta=0}\left[\left(1-x\right)N_{q}^{AB}+ x N_{q}^{AB}\nu_{q}^{AB}\right],
\end{equation}
where $(\frac{dn_{ch}}{d\eta})^{pp}_{\eta=0}$ is calculated from Eq. (2) using the new parameter values.
Here $x$ signifies the relative contributions of hard and soft processes in two component model ~\cite{Thomos:2009,Thomos:2011}. However we found in our earlier publication ~\cite{Chaturvedi:2016} that the midrapidity particle density properly scales with number of wounded quarks at RHIC and LHC (Large Hadron Collider) and thus supports the idea of one component WQM in central rapidity region. In present calculation, we have again found that the value of $x$ is negligibly small if our WQM has to satisfy the minimum bias $U$-$U$ experimental data at $193$ GeV. Recent experimental data of charged hadrons in $U$-$U$ collisions ~\cite{Adam:PRL2015} show a small correlation between the multiplicity or elliptic flow of charged hadrons and the initial geometrical orientations. However Monte Carlo Glauber model (MCGM) shows contradictory behavior when using two component model ~\cite{Sandeep:2016,Moreland:2015}. It has been suggested that this contradiction arises due to contribution of collision term in the two-component wounded nucleon model. In the view of this we have taken a very small value of $x$ as $0.005$ to minimize the contribution of collision term in our calculations and try to see whether our model satisfy the minimum bias experimental data and if yes then whether it shows a small or large correlation between multiplicity and the initial orientations of $U$-$U$ collisions.

We further extend our two component formula of pseudorapidity distribution to provide it a $\eta$-dependence using a new parametrization which is based on Landau's distribution function for it as follows :

\begin{equation}
\left(\frac{dn_{ch}}{d\eta}\right)^{AB} = 2 \left(\frac{dn_{ch}}{d\eta}\right)^{AB}_{\eta=0}\times \frac{\sqrt{1-\frac{1}{(\beta cosh\eta)^{2}}}}{\gamma+exp(\eta^{2}/2\sigma^{2})},
\end{equation}
where $\beta,~\gamma$ and $\sigma$ are fitting parameters, and $(\frac{dn_{ch}}{d\eta})^{AB}_{\eta=0}$ is the calculated from Eq. (11). Here one should keep in mind that this parametrization is able to give the pseudorapidity distribution of charged particles produced in symmetric collisions (like body-body or tip-tip but not body-tip configuration) of nuclei. We will show how these parameters are related with the shape of distribution in different limits.
\\
{\bf Case I :} when $\eta$ is very large then
\begin{eqnarray}
cosh\eta &=& \infty \\ \nonumber
\implies \sqrt{1-\frac{1}{(\beta cosh\eta)^{2}}}&=&\sqrt{1-\frac{1}{\infty}}=1
\end{eqnarray}
Now comparing the original Landau distribution with our distribution in large $\eta$ limit :
\begin{eqnarray}
\frac{1}{\gamma+exp(\eta/2\sigma^{2})}=exp(-\eta/2\sigma^{2})
\end{eqnarray}
Multiplying both side of equation with $1/exp(\eta/2\sigma^{2})$, we get :
\begin{equation}
\gamma exp(-\eta^{2}/2\sigma^{2})=0
\end{equation}
From here we can see that if $\gamma$ is zero then our distribution completely takes the form of Landau distribution. It means in large $\eta$ limit, only the $\gamma$ parameter which differ our new distribution function  from the original Landau distribution function for charged hadron production at forward and backward rapidity.\\
{\bf Case II :} when $\eta$ is equal to zero then
\begin{equation}
\sqrt{1-\frac{1}{(\beta cosh\eta)^{2}}}=\sqrt{1-\frac{1}{\beta^{2}}}
\end{equation}
Now comparing the numerator of our distribution to the numerator of Landau distribution in $\eta=0$ limit:
\begin{eqnarray}
\sqrt{\beta^{2}-\frac{1}{\beta^{2}}}&=&1 \\ \nonumber
\implies \frac{1}{\beta^{2}}&=&0
\end{eqnarray}
So the numerator of our distribution in $\eta=0$; if we take $\beta$ is much larger than $1$ :
\begin{equation}
\gamma exp(-\eta^{2}/2\sigma^{2})=0
\end{equation}
Thus in $\eta=0$ limit, the parameter $\beta$ should be large and $\gamma$ must be zero then only our distribution will take the similar structure as Landau's distribution. 

In this article we also calculate the transverse energy density distribution of charged hadrons using the pseudorapidity distribution of WQM as follows :
\begin{equation}
dE_{T}/d\eta \cong \frac{3}{2} \sqrt{{\left\langle p_{T} \right \rangle}^2+m_{\pi}^2} (dn_{ch}/d\eta),
\end{equation}
where $\langle p_{T}\rangle$ is the average transverse momentum of the produced charged particles and $m_{\pi}$ is the mass of pion. Further we have calculated the Bjorken's energy density~\cite{bjorken} in $U$-$U$ collisions using following formula :
\begin{equation}
\varepsilon_{BJ} \cong \frac{3}{2} \sqrt{{\left\langle p_{T} \right \rangle}^2+m_{\pi}^2}(dn_{ch}/d\eta)/ \tau \pi R^{2}.
\end{equation}
In the above expression, $\tau$ is the hadronic formation time. 
\section{Results and Discussions}
\subsection{Total Multiplicity and Pseudorapidity density}
\begin{figure}
\includegraphics[scale=0.50]{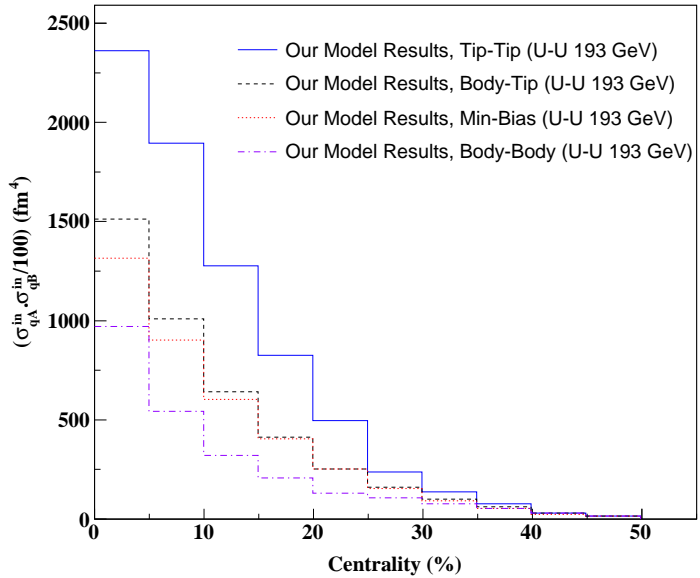}
\caption{(Color online) Variation of the quark-nucleus inelastic cross-section ($\sigma_{qA}^{in}$) in our model as a function of centrality for various initial configurations of $U$-$U$ collisions at $\sqrt{s_{NN}}$ = $193$ GeV. }
\end{figure}

\begin{figure}
\includegraphics[scale=0.50]{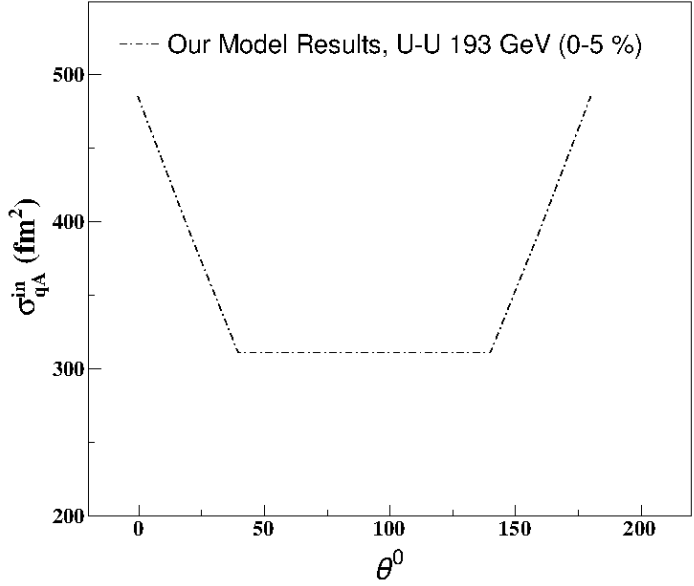}
\caption{Variation of $\sigma_{qA}^{in}$ in our model as a function of polar angle $\theta$ for central $U$-$U$ collisions.}
\end{figure}

\begin{figure}
\includegraphics[scale=0.50]{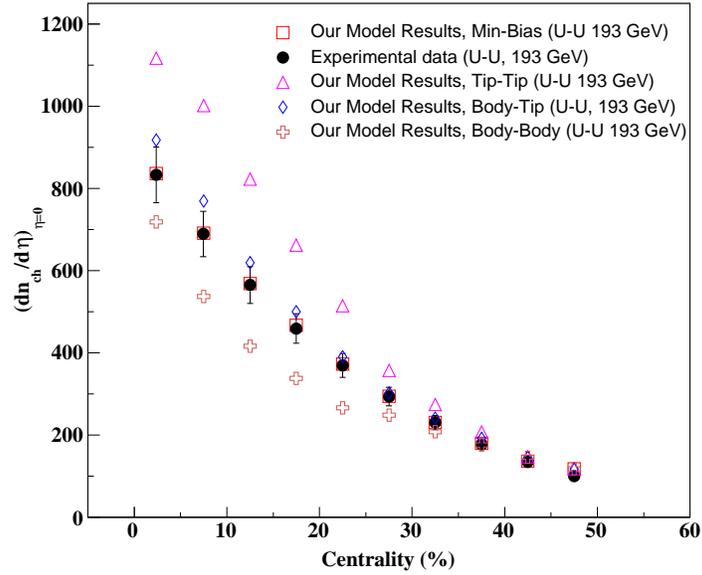}
\caption{(Color online) Variation of pseudorapidity density of charged hadrons produced in $U$-$U$ collisions with respect to centrality for various initial geometrical configurations. Experimental data with respect to centrality is also shown here for min-bias (contribution from all the initial configuration of $U$-$U$ collisions)~\cite{Adare:2016}.}
\end{figure}

\begin{figure}
\includegraphics[scale=0.50]{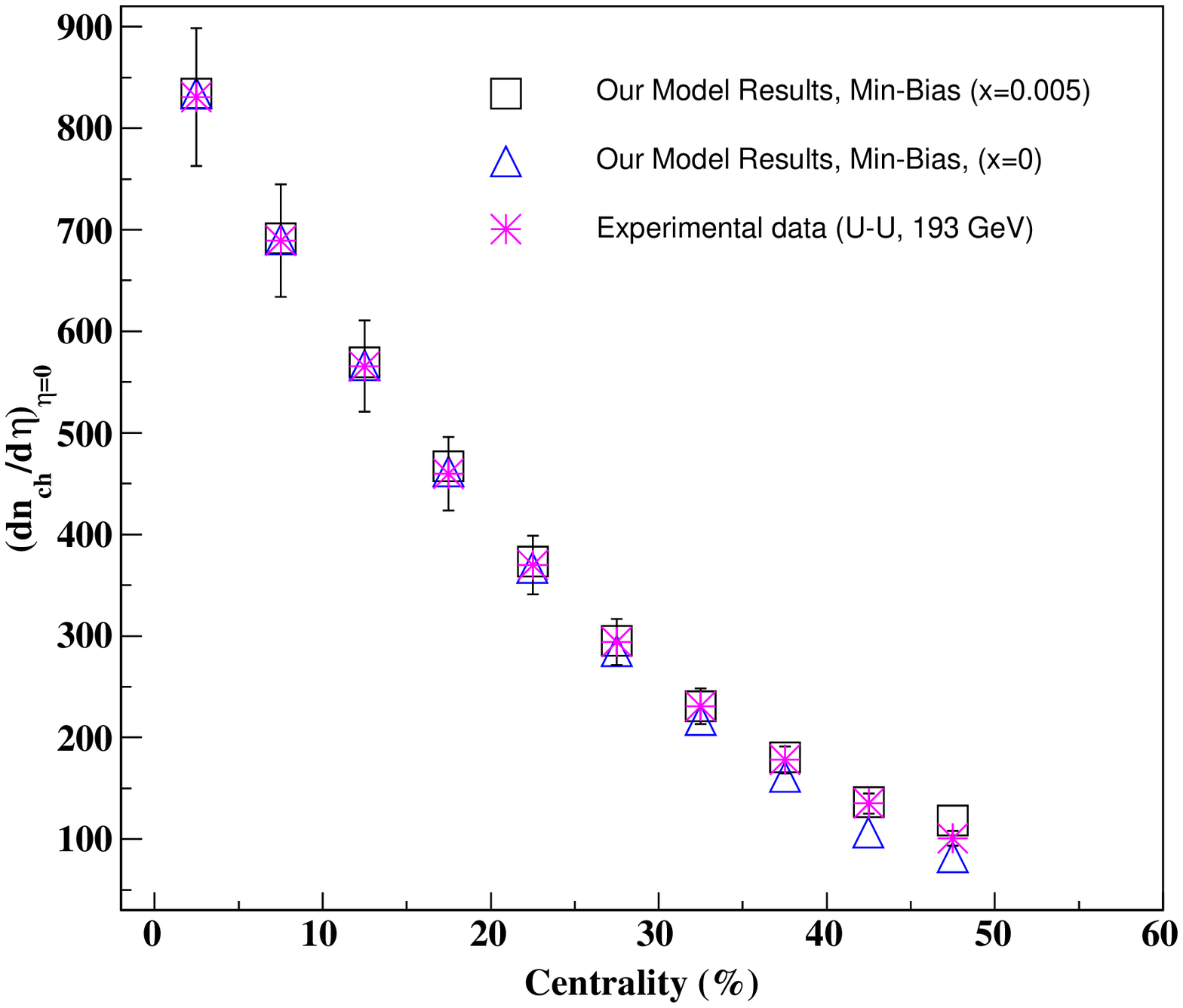}
\caption{(Color online) Variation of pseudorapidity density of charged hadrons produced in $U$-$U$ collisions with respect to centrality for min-bias configuration. We have shown the results for $x=0.005$ and $x=0$, where $x$ controls relative contribution of hard process. Experimental data is taken from Ref.~\cite{Adare:2016}.}
\end{figure}

In our earlier publications we have shown that WQM provides proper and reasonable description to the various features of charged hadron production in high-energy collisions for symmetric (e.g., $Au$-$Au$,~$Pb$-$Pb$, $Cu$-$Cu$ etc.) as well as for asymmetric (e.g., $Cu$-$Au$,~$d$-$Au$ etc.) collisions. Here we will study the different properties of charged hadron production in the collisions of deformed uranium nuclei at $\sqrt{s_{NN}}$ = $193$ GeV~\cite{Adare:2016}. Fig. 3 presents the variation of $\sigma_{qA}^{in}\cdot\sigma_{qB}^{in}$ with respect to centrality for various initial configurations of $U$-$U$ collisions at $\sqrt{s_{NN}}$ = $193$ GeV~\cite{Adare:2016}. From figure one can observe that the effect of various configurations of $U$-$U$ collision is negligible for most peripheral collisions. However, in central collision the quark-nucleus scattering cross-section for tip-tip is considerably high in comparison to other configurations. This is due to a large number of color strings formed between the participating quarks of nucleus $A$ as well as $B$ in tip-tip configuration. In body-tip geometry, the participating quarks within nucleus $A$ is less than the participating quarks within nucleus $B$. Thus the total number of color strings formed in body-tip collision is smaller than tip-tip but larger than body-body collisions. These numbers actually reflects in the scattering cross-sections of various initial configurations. The product of $\sigma_{qA}^{in}$ and $\sigma_{qB}^{in}$ for minimum bias (in terms of configurations not in terms of centrality) matches with the body-tip configurations over the entire centrality range. In Fig. 4, we have shown the change in quark-nucleus scattering cross-section with respect to polar angle $\theta$ of nuclear charge density distribution function in central $U$-$U$ collisions. It varies from its maximum value at $\theta=0$ and $180^{0}$ to a minimum value at $90^{0}$. The change in $\sigma_{qA}$ values is about $40\%$ depending on the polar angle. We have also checked the change in value of $\sigma_{qA}$ by using modified value of parameter as given in Ref.~\cite{Shou:2015}. The change is less than $1\%$ thus we stick to our earlier values of parameters.

\begin{figure}
\includegraphics[scale=0.50]{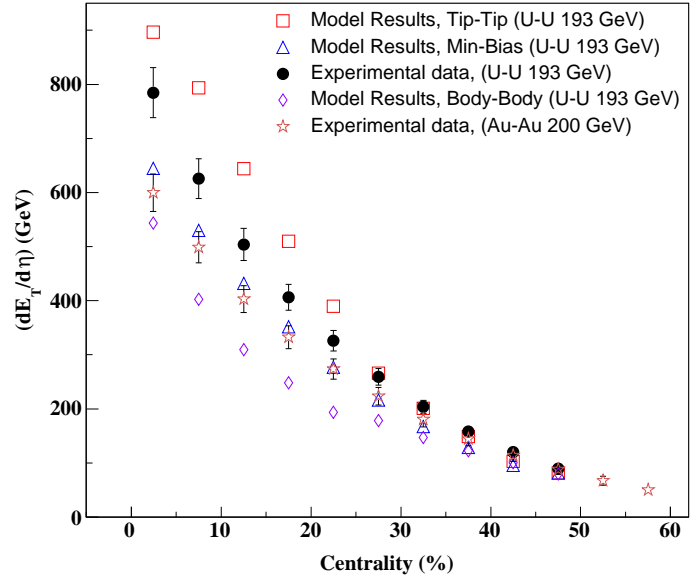}
\caption{(Color online) The transverse energy density distribution of charged hadrons produced in $U$-$U$ collisions~\cite{Adare:2016} is shown with centrality for tip-tip, body-body and body-tip configurations.}
\end{figure}

\begin{figure}
\includegraphics[scale=0.50]{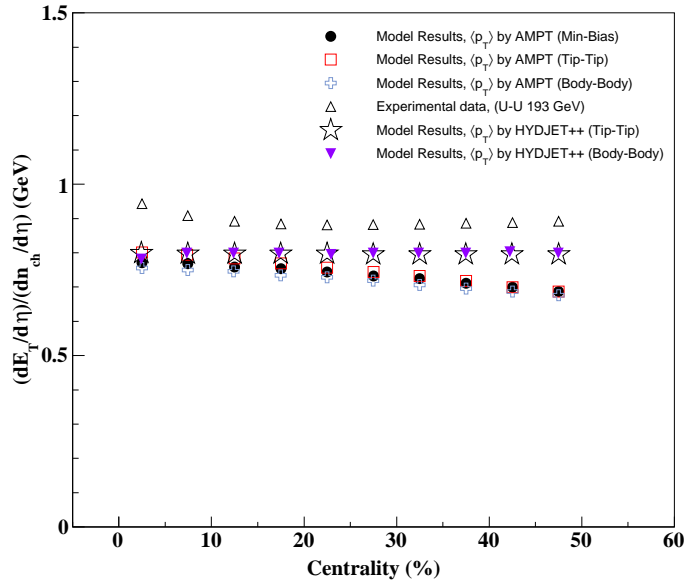}
\caption{(Color online) Ratio of $(dE_{T}/d\eta) / (dn_{ch}/d\eta)$ at mid-rapidity is shown in our model using $\langle p_{T}\rangle$ from AMPT model~\cite{Rihan:2012} as well as from HYDJET++ model~\cite{arpit}. Comparison with experimental data ~\cite{Adare:2016} is also shown.}
\end{figure}

\begin{table*}
\begin{center}
\caption{The obtained values of $N_{q}$ by Eq. (9) in $U$-$U$ at $193$ GeV for its different geometrical configurations. $N_{q}$ for $p$-$p$ collision is equal to $1$ which is also mentioned in AQM~\cite{BialasPRD:1982,Anisovich:1984,lipkin} for hadron-hadron collisions.}
\begin{tabular}{|l|l|l|l|l|l|l|l|l|l|l|l|l}
\hline
{Centrality Bin} & \multicolumn{4}{c|}{ Model Calculation }   \\
\cline{2-5}
 &{Tip-Tip}&{Body-Tip}&{Min-Bias}&{Body-Body} \\
\hline\hline

  ~$0-5 \%$  & 444.28     & 364.18  & 331.12  & 284.09            \\
 ~$5-10 \%$  & 398.45     & 305.22  & 274.03  & 211.98       \\
$10-15 \%$   & 326.64     & 244.98  & 224.74  & 163.32         \\
$15-20 \%$   & 262.36     & 196.77  & 183.42  & 131.18           \\
$20-25 \%$   & 202.99     & 152.25  & 145.25  & 101.49              \\
$25-30 \%$   & 139.21     & 116.69  & 113.11  & 94.17              \\
$30-35 \%$   & 104.91     & 90.97   & 86.17   & 77.02             \\
$35-40 \%$   & 76.66     &  69.78   & 63.97   & 62.89               \\ 
$40-45 \%$   & 48.92     &  48.76   & 42.11   & 41.61           \\
$45-50 \%$   & 32.66     &  32.66   & 32.66   & 32.66            \\ \hline
             
\end{tabular}
\end{center}
\end{table*}

\begin{figure}
\includegraphics[scale=0.50]{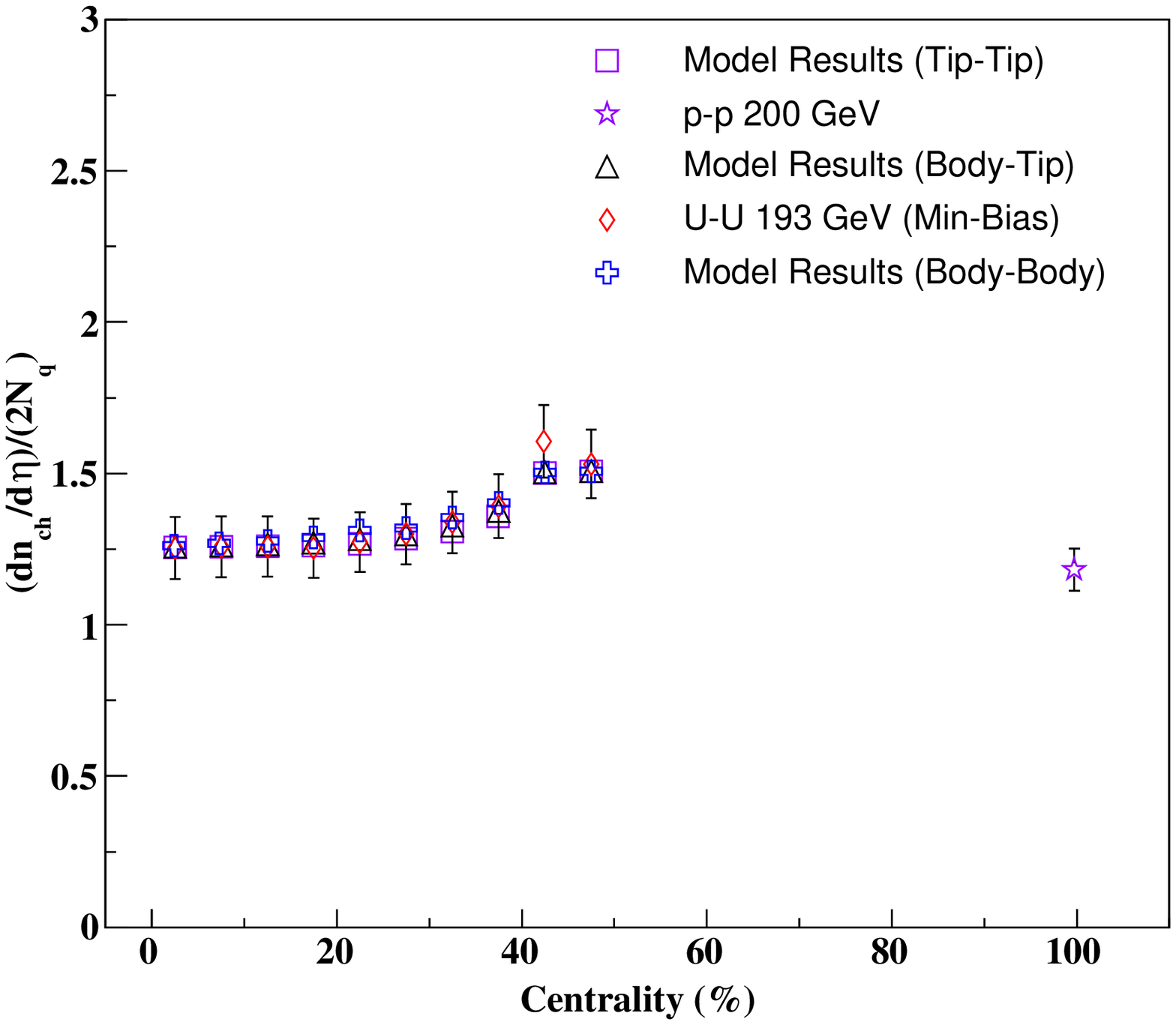}
\caption{(Color online) Variation of $(dn_{ch}/d\eta)/(2 N_{q})$ as a function of centrality for $U$-$U$ collisions at $\sqrt{s_{NN}}$ = $193$ GeV. We have also shown $(dn_{ch}/d\eta)/(2 N_{q})$ for $p$-$p$ collisions at $\sqrt{s_{NN}}$ = $200$ GeV~\cite{Nouicer}. Here $N_{q}$ is calculated using Eq. (9) and experimental data of $(dn_{ch}/d\eta)$ is taken from Refs.~\cite{Adare:2016,Nouicer}.}
\end{figure}

The main parameters by which one can select the different initial configurations of $U$-$U$ collisions are pseudorapidity distribution ($dn_{ch}/d\eta$) and transverse energy distribution ($dE_{T}/d\eta$) of charge hadrons at midrapidity. Fig. 5, demonstrates the variation of $(dn_{ch}/d\eta)_{\eta=0}$ with centrality for different geometrical configuration of $U$-$U$ collisions. Our model results (we have used $x=0.005$ in calculating $(dn_{ch}/d\eta)_{\eta=0}$ from Eq. (11)) for minimum bias configuration suitably matches with the experimental data of the same configuration for all centrality intervals. The multiplicity is larger for tip-tip and body-tip configuration in central and semi-central collisions due to higher compression and a longer passage time of the reaction. The particle multiplicities for tip-tip configuration in most central collision is around $36\%$ higher than body-body and $27\%$ higher than minimum bias. However, it is not possible to differentiate among various configurations of $U$-$U$ collisions in peripheral and semi-peripheral collision as far as $dn_{ch}/d\eta$ is concerned. One should apply some other variable cut along with $dn_{ch}/d\eta$ to select desired configuration in peripheral and semi-peripheral events. Body-body configuration have least multiplicity of charged hadron in central and semi-central collisions. In Fig. 6, we have shown $(dn_{ch}/d\eta)_{\eta =0}$ for min-bias configuration for two different values of $x=0.005$ and $0$, where $x$ signifies the relative contribution of hard and soft process. From figure it is clear that we need a little contribution of hard processes in semi-peripheral and peripheral events to properly satisfy the experimental data. However, in central events the contribution of hard process remains negligible for both the values of $x$.   

Recent experimental data on $U$-$U$ collisions ~\cite{Adam:PRL2015} have demonstrated a small correlation between multiplicity and the initial orientation of uranium nuclei. However our model results show a different behavior. One possibility is lack of any proper way to disentangle the various configurations of $U$-$U$ collisions experimentally. Thus we have to study the other observables too along with $dn_{ch}/d\eta$. In this regard, we have calculated another observable $dE_{T}/d\eta$ which is a quantitative measure of the energy deposited mainly by the produced particles (not the fragments) in the transverse plane of the collision and is related with the explosiveness of the collision~\cite{skt2}. In Fig. 7, we have shown the variation of $(dE_{T}/d\eta)_{\eta=0}$ in different centrality bins for possible configurations. It is important here to note that the average transverse momentum ($\langle p_{T}\rangle$) used in the calculation of transverse energy density distribution is taken from Ref.~\cite{Rihan:2012} in which author calculated $\langle p_{T}\rangle$ in $U$-$U$ collision at $200$ GeV by AMPT model (a multi-phase transport model)~\cite{Rihan:2012}. Here our min-bias result does not describe the experimental data suitably for central and semi-central events. However the trend of $dE_{T}/d\eta$ is similar to $dn_{ch}/d\eta$ for various configuration at all centralities.

\begin{figure}
\includegraphics[scale=0.50]{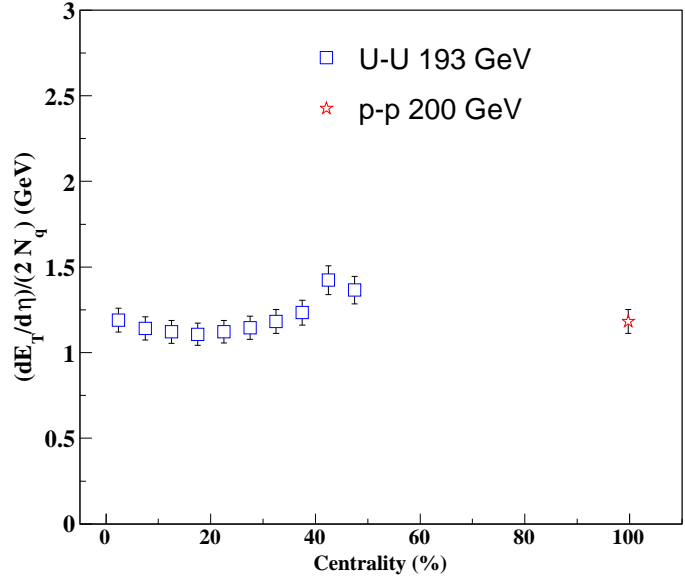}
\caption{(Color online) Variation of $(dE_{T}/d\eta)/(2 N_{q})$ as a function of centrality for $U$-$U$ collisions at $\sqrt{s_{NN}}$ = $193$ GeV. We have also shown $(dE_{T}/d\eta)/(2 N_{q})$ for $p$-$p$ collisions at $\sqrt{s_{NN}}$ = $200$ GeV~\cite{Nouicer,Alver}. Experimental data of $(dE_{T}/d\eta)$ is taken from Ref.~\cite{Adare:2016}.}
\end{figure}

\begin{figure}
\includegraphics[scale=0.50]{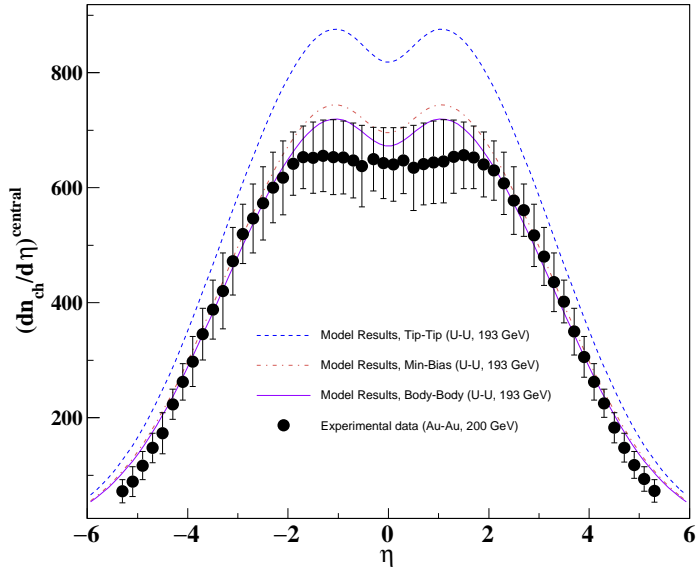}
\caption{(Color online) Pseudorapidity distribution of charged hadrons with respect to pseudorapidity for various configuration of $U$-$U$ collisions.}
\end{figure}

\begin{figure}
\includegraphics[scale=0.50]{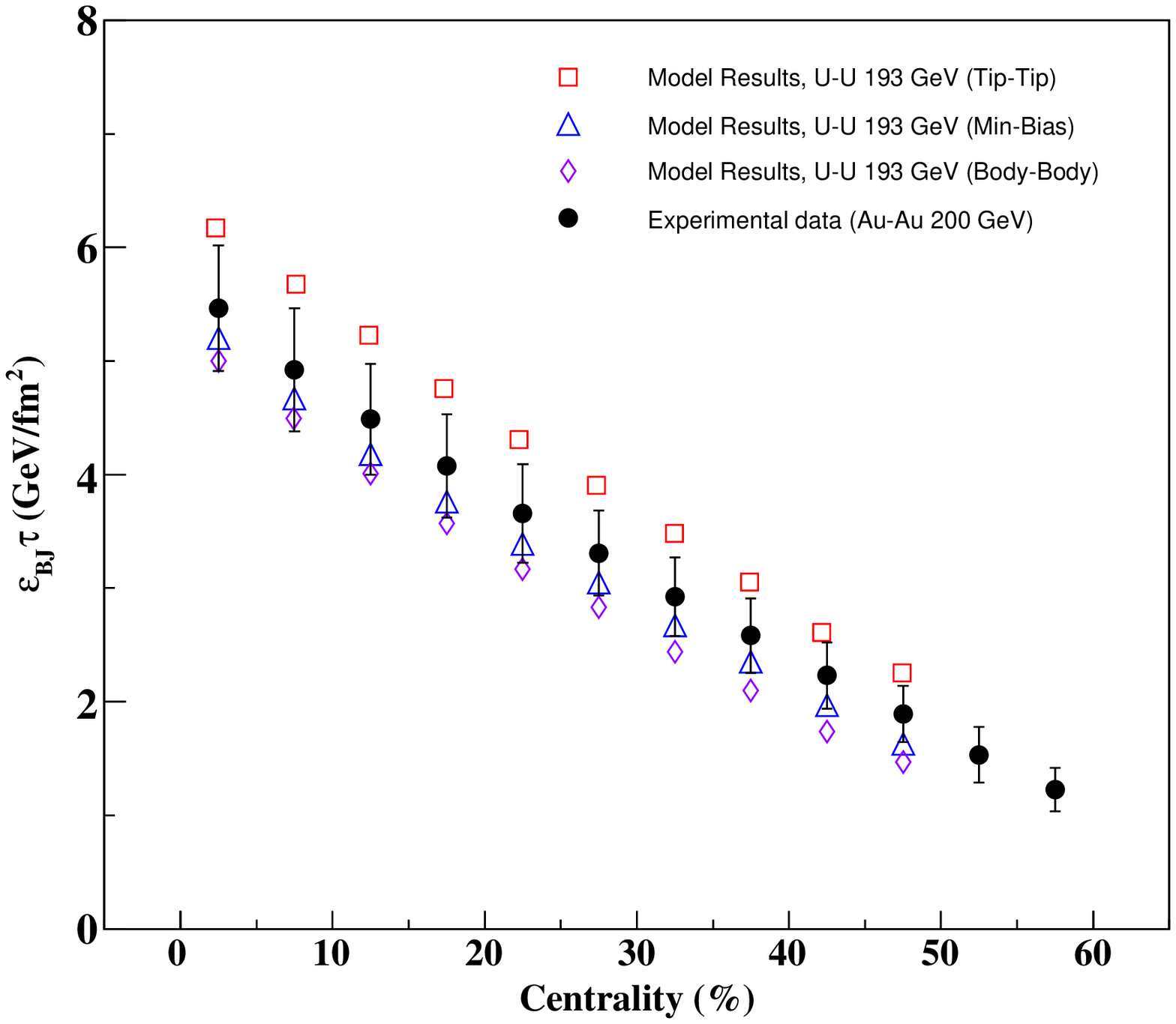}
\caption{(Color online) Variation of product of Bjorken's energy density and proper time with respect to centrality for  various configuration of $U$-$U$ collisions.}
\end{figure}

The ratio $(dE_{T}/d\eta)/(dn_{ch}/d\eta)= dE_{T}/dn_{ch}$ provides the value of mean transverse energy per produced hadron ($E_{T}/N_{ch}$) and shed light on the possible particle production mechanism and the freeze-out criteria~\cite{skt2,skt1}. It has been observed that the value of this ratio is almost $1$ starting from low energy AGS (Alternating Gradient Synchrotron) experiment upto highest possible energy at RHIC. In Fig. 8, we have plotted this ratio in our model as well as using experimental data for various possible configuration with respect to centrality in $U$-$U$ collisions at $\sqrt{s_{NN}}=193$ GeV. As we have mentioned earlier that we have used $\langle p_{T}\rangle$ from AMPT model in calculation of $dE_{T}/d\eta$ and thus we have shown the ratio $(dE_{T}/d\eta)/(dn_{ch}/d\eta)$ using AMPT model. Further, we have also shown $(dE_{T}/d\eta)/(dn_{ch}/d\eta)$ using HYDJET++ (Hydrodynamics plus JETs) model for tip-tip and body-body configurations in which we have used $\langle p_{T}\rangle$ from HYDJET++ model in calculation of $dE_{T}/d\eta$~\cite{arpit}. From figure one can see that the ratio is almost constant and has a value near $0.9$ in experimental data of min-bias. The ratio varies between $0.8$ to $0.85$ in HYDJET++ case and from $0.8$ to $0.7$ in AMPT case for every configuration. The difference between experimental and model results may be due to the uncertainty in calculating $dE_{T}/d\eta$ using average transverse momentum from various models~\cite{Rihan:2012,arpit}. Independence of this ratio ($E_{T}/N_{ch}$) from centrality and geometrical orientations supports the idea of chemical freezeout and particle production from a single freezeout surface. The slight dependence over centrality can be seen which possibly hints the contribution of collective dynamics on particle production.

Fig. 9 presents the variation of ratio $(dn_{ch}/d\eta)/(2 N_{q})$ as a function of centrality for $U$-$U$ collisions at $\sqrt{s_{NN}}$ = $193$ GeV in various configurations. In minimum bias case, we use experimental data of $dn_{ch}/d\eta$ at mid-rapidity and divide it by the number of participating quark $N_{q}$ calculated in our WQM (as shown in Table 1) for minimum bias configuration. We found that the ratio in this case shows a scaling pattern with the experimental data of $dn_{ch}/d\eta$ in $p$-$p$ collision at $200$ GeV~\cite{Nouicer,Alver} divided by the number of participating quarks. Only a slight violation of scaling is found in peripheral collisions. For other configuration, as the experimental data of $dn_{ch}/d\eta$ is unavailable thus we use the model results for both $dn_{ch}/d\eta$ and $N_{q}$ to show the scaling pattern. In Fig. 10, we have found the same scaling pattern for minimum bias $dE_{T}/d\eta$ experimental data with $N_{q}$ and compare it with the corresponding data in $p$-$p$ collision at $200$ GeV~\cite{Nouicer,Alver}. These results suggest that the quark-quark interactions is more suitable in comparison to nucleon-nucleon interactions in describing the nucleus-nucleus collisions. 

In Fig. 11, we have calculated the pseudorapidity distribution of charged hadrons in central $U$-$U$ collisions and demonstrate its variation with $\eta$ for various geometrical orientations.  We have also compared our WQM results for $U$-$U$ collisions with the corresponding experimental data in $Au$-$Au$ collisions at $200$ GeV~\cite{Back_PRC}. Here we observe that the charged hadron production in tip-tip $U$-$U$ central collisions is about $1.5$ times larger than the $Au$-$Au$ collision. Even the charged hadron multiplicity at mid-rapidity in body-body and minimum bias orientations is larger in comparison to particle produced in $Au$-$Au$ collisions. This suggests that the effect of various configurations of deformed uranium nucleus is more on the produced particles rather than fragments. The present method of calculating the pseudorapidity distribution with $\eta$ is inapplicable for body-tip configurations due to their asymmetric distribution in forward and backward rapidity. There are some efforts to show the pseudorapidity distribution in body-tip configuration~\cite{body_tip}.

In Fig. 12, we have shown the product of Bjorken's energy density ($\epsilon_{BJ}$) and hadronic formation time ($\tau$) with respect to centrality for various orientational configuration of $U$-$U$ collisions. We have compared our model results with the experimental data in $Au$-$Au$ collision at $200$ GeV~\cite{Adare:2016} to show the increase in initial energy density of the fireball formed in $U$-$U$ collisions. In central collisions, tip-tip configuration of uranium nuclei produces the most dense QCD system at this much energy which is about $30\%$ higher than the energy density of the fireball created in $Au$-$Au$ collision. The energy density produced in minimum bias configuration is approximately equal to the corresponding quantity in $Au$-$Au$ collisions within error bars. Body-body configuration of central $U$-$U$ collision creates a system having least energy density among all the cases considered in this figure. 
\section{Summary and Conclusions}
In summary, we have shown the available energy dependence of all the terms (constant term, linear term and quadratic term) of the parametrization used by us to calculate pseudorapidity density in $p$-$p$ collisions. After that we have shown the schematics of special geometrical orientation of $U$-$U$ nuclei in different configurations viz. tip-tip, body-tip and body-body. To set different centrality class in $U$-$U$ collisions, we have plotted the inelastic cross section $\sigma_{qA}$ for tip-tip, body-tip, min-bias and body-body configurations and found that there is a systematic variation with  centrality in different configurations. The variation of $\sigma_{qA}$ as a function of polar angle is also shown. We have then estimated the pseudorapidity density at mid-rapidity for different configurations in $U$-$U$ at $\sqrt{s_{NN}}$ = $193$ GeV within modified WQM and compared with the RHIC experimental data in min-bias and found a good agreement in this configuration. The variation of transverse energy density with centrality for min-bias, tip-tip, and body-body configurations are also evaluated using pseudorapidity distributions which we have obtained in WQM. We observe an enhancement in multiplicity for tip-tip collisions for most central events compared to experimental multiplicity in $Au$-$Au$ collisions at RHIC energies. We have shown the ratio of $E_{T}/N_{ch}$ ($\cong{dE_{T}/d\eta/dn_{ch}/d\eta}$) for tip-tip, min-bias and body-body configurations using WQM as well as experimental data and observed that the dependence of this ratio on centrality is not significant. We have plotted the $dn_{ch}/d\eta/(2N_{q})$ and $dE_{T}/d\eta/(2N_{q})$ as a function of centrality, where $N_{q}$ is evaluated within wounded quark scenario, for different configurations and  observed a scaling behavior in the production of charged hadrons. We have predicted the detailed description of $\eta$ variation of pseudorapidity density within our modified Landau description and discuss the effect of this modification in different pseudorapidity regime. The effect of various configurations is also presented in the plots. Moreover we have given the prediction of Bjorken energy density as a function of centrality in $U$-$U$ collisions for different geometrical configurations and compared the result with the energy density produced in $Au$-$Au$ collision experiment at highest RHIC energy. 
Finally we conclude that the present version of WQM suitably describes the experimental findings and also highlights the role of quark-quark interactions in the production of charged hadrons in ultra relativistic heavy-ion experiments.

\section{Acknowledgments}
\noindent 
O S K Chaturvedi is grateful to Council of Scientific and Industrial Research (CSIR), New Delhi for providing a research grant. PKS acknowledges IIT Ropar, India for providing an institute postdoctoral research grant. We would like to thank authors of Ref.~\cite{arpit} for providing the $\langle p_{T}\rangle$ values by HYDJET++.

\end{document}